\documentclass[twocolumn,10pt]{asme2ej}
\usepackage{epsfig} 
\graphicspath{{figures/}} 
\usepackage{graphicx}
\usepackage[utf8]{inputenc}
\usepackage{lmodern}
\usepackage[figurename=Fig.,labelfont=bf,labelsep=period]{caption}
\usepackage{siunitx}
\usepackage{subcaption}
\usepackage{amsmath}
\usepackage{newtxtext,newtxmath}
\usepackage[colorlinks=true,citecolor=red,linkcolor=black]{hyperref}
\usepackage[capitalise]{cleveref}
\usepackage{bm}
\usepackage{mathtools,upgreek}
\usepackage{tabularx,ragged2e,booktabs,caption}
\usepackage{bigdelim,bigstrut}
\usepackage{chngcntr}
\usepackage[square,numbers]{natbib}
\renewcommand{\vec}[1]{\bm{\mathrm{#1}}}
\newcommand{\etal}{\textit{et~al}. }
\hypersetup{
	colorlinks=true,
	linkcolor=blue,
	filecolor=blue,      
	urlcolor=blue,
}
\newcommand{\im}{\mathrm{i}}
\newcommand{\bLozenge}{\mathbin{\blacklozenge}}

\title{Reduced Order Modeling of Dynamic Mechanical Metamaterials for Analysis of Infinite and Finite Systems}

\author{Weidi Wang
 \affiliation{
 Ph.D. Candidate\\
 Department of Mechanical Engineering\\
 University of Massachusetts, Lowell\\
 Lowell, Massachusetts 01854, USA}
 }
 
\author{Alireza V. Amirkhizi
  \affiliation{Associate Professor\\
  Department of Mechanical Engineering\\
  University of Massachusetts, Lowell\\
  Lowell, Massachusetts 01854, USA\\
  Email: alireza\_amirkhizi@uml.edu}
  }

\begin{document}
\maketitle

\begin{abstract}{
Dynamic mechanical metamaterials (MMs) are artificial media composed of periodic micro-structures, designed to manipulate wave propagation. Modeling and designing MMs can be computationally demanding due to the broad design space spanned by the geometric and material parameters. This work aims to develop a generalized reduced order modeling (ROM) approach for determining MM dynamics in low frequency ranges with accuracy and speed, using a limited number of parameters and small matrices. The MM unit cells are treated as assemblies of structural elements with discrete degrees of freedom, whose effective stiffness and inertia are determined by optimizing energy criteria based on continuum results derived from a small number of eigen-study simulations. This proposed approach offers a parameterized and discretized representation of MM systems, which leads to fast and accurate computation of eigen-study results for periodic arrays, as well as dynamic responses in time domain for finite-sized arrays. The high computational efficiency and physical accuracy of this method will help to streamline the modeling process and aid in design discovery and optimization, especially in combination with machine learning and data-driven techniques.
}
\end{abstract}

\section{Introduction}

Dynamic mechanical metamaterials (MMs) feature sub-wavelength micro-structures that interact with stress waves, exhibiting exotic functionalities.
Numerous exciting potentials have been proposed for MMs, including wave attenuation \cite{Chen2017f,Fang2021dd,Baertsch2021}, negative refraction \cite{Ding2010,Seo2012,Li2016,Nemat-Nasser2015a}, cloaking \cite{Chen2007,Norris2011,Zhu2014,Cummer2016}, and insulators \cite{Oh2017,Matlack2018}.
The systematic design of metamaterials requires a comprehensive understanding of the dynamic behaviors of MM itself as well as manufacturing design constraints. The first natural step is to find the eigenfrequency band structure and mode shapes of the design. The former encodes major characteristic information of a MM unit cell such as resonance frequencies, wave velocities, and band gaps, while the latter is needed to determine scattering in finite structures and to evaluate interactions between modes fully. 
Common methodologies of band structure calculation include plane wave expansion (PWE) \cite{Wu2004g,Sridhar2017,Lu2017a,Oudich2014}, transfer matrix method (TMM) \cite{Mead1996,Junyi2015,Amirkhizi2018c}, and finite element method (FEM) \cite{Huang2021}.
In almost all cases, the design and analysis of these dynamic systems are plagued by geometric complexity and computational burden. The major challenges include 1) unclear design-performance relationship and 2) expensive computational cost for calculating the dynamic response. 
Therefore, for design optimization purposes, a reduced order modeling (ROM) method is clearly more suitable as it allows for simple and fast computation limited to frequency range or modalities of interest.

In this paper we present a ROM approach for fast computation of MM problems, which can be used for different study setups, including eigenfrequency band calculation and computation of time dependent dynamic responses of finite structures.
The metamaterials that are considered here are assumed to be comprised of 2D micro-structural designs with beam-like elements. The materials are assumed to have no loss or gain mechanisms, but the inclusion of linear viscoelastic response can be considered as a natural future expansion. Detailed numerical and experimental studies on some similar micro-structures can be found in previous work \cite{Aghighi2019,Amirkhizi2018d}.

Extensive reduced order modeling techniques have been developed for vibration problems, e.g. dynamic condensation \cite{Kidder1973}, improved reduced system (IRS) \cite{Gordis1994}, and system equivalent reduction expansion process (SEREP) \cite{Avitable1989}. These reduction methods in general employ certain transformation matrices that map the full set of degrees of freedom (DOFs) to a reduced set of DOFs.
For wave propagation problems, especially metamaterial problems, the existing model order reduction methods are limited and less applicable because the system matrices and the eigenfunctions are dependent on the wavevector. 
The wavevector-dependence leads to the frequency and mode variation in the band structures and is a key element in metamaterial dynamics. Therefore, novel reduction schemes that can preserve the wavevector dependence and band accuracy are needed for metamaterial problems. To this end, Hussein \cite{Hussein2009a} introduced reduced Bloch mode expansion (RBME) for fast computation of band structures. The RBME method employs selected Bloch eigenfunctions to reduce the dimensionality. 
A similar method, Bloch mode synthesis (BMS) \cite{Krattiger2014,Krattiger2018a}, is an extended sub-structuring technique that describes the structural DOFs by normal and constraint modes. Both the RBME and BMS methods utilize selected eigen-modes to construct transformation matrices that reduce the size of the full matrices. These transformation-based methods effectively reduce the number of equations, but the resulting matrices are no longer representing the physical quantities (stiffness and inertia), therefore less suitable for geometric or material design problems. Additionally, these methods could not be applied to time/frequency domain computations of finite arrays. Nevertheless, these methods have been shown useful for topology optimization \cite{Jung2020} in terms of reducing the computational cost.
An alternative scheme is to develop discrete models comprised of masses and springs. 
The discretized mass-spring representation has been widely accepted in the literature as it offers analytical formulations that simplify the computational effort while retaining essential physics. It has proven to be beneficial for various design aspects such as feasibility analysis \cite{Dertimanis2016}, reliability assessment \cite{Wagner2018}, and design space mapping \cite{Morris2022}. For higher order systems operating at high frequency ranges, an excellent example of modeling discrete weakly coupled MMs has been introduced by Matlack~\etal \cite{Matlack2018}, where the model reduction is performed using the Schrieffer-Wolff transformation so that the modes in the frequency range of interest are decoupled. 
However, this method could only be applied to narrow-band dynamics.

While the mass-spring representation can significantly reduce the computational effort, certain vibration modes may exhibit mixed coupling between DOFs. To accurately capture such dynamics, the elastic spring elements cannot be simple 2-DOF elements.
In our previous work \cite{Amirkhizi2018d}, the elastic spring elements are physically represented as beams. The reduced stiffness and mass matrices can then be derived using simple strength of materials analysis. Such an approach provides analytical matrices that operate on physical DOFs and is naturally suitable for tuning the response via control of physical dimensions and material choices \cite{Morris2022}. They also make interpreting the modal physics straightforward, for example, such a beam-based discrete model allows for accurate identification of the level repulsion \cite{Lu2018} and coupling between the DOFs. However, the selection of DOFs has been mostly a heuristic step that may affect the results. 
In addition, approximating the structural components as standard beam elements does not generally match the actual response of the beam-like elements as accurately as needed.

The present work introduces a systematic implementation of the structural-element-based ROM approach that overcomes these limitations. 
A generalized ROM procedure, parameterized in terms of effective structural stiffness parameters and discrete DOF inertia, is developed and is shown to be applicable to a large family of 3D printable MM designs. 
The conceptual idea of the proposed ROM method takes advantage of the fact that the 3D printable MMs that operate as low frequency (long wavelength) locally resonant systems are often comprised of slender plate- or beam-like elements. 
In addition, in most cases only the low frequency dynamics of the MM are of particular interest for practical applications, which reside in the subspace spanned by the lowest few eigen-modes, representable using a few carefully selected DOFs.
Modeling the system with these ``master" DOFs can hence reduce the computational cost while maintaining high fidelity of the underlying physics.
A structural assembly system with symbolic matrices is used to represent the repeating unit cell (RUC). By optimizing the energy fitness compared with numerical results, one can find the effective stiffness and inertia parameters of this structural assembly. Such a ROM unit cell can accurately predict the eigenfrequency band structures with minimal computational effort.
This approach improves upon existing model order reduction methods in handling problems in metamaterials by maintaining eigen-solution accuracy within the Brillouin zone and providing parameterized matrices. It incorporates the propagating nature of waves, rather than just the modal response of finite structures. In MM systems, small variations in geometry can drastically change the overall response. Using an analytical model that characterizes the MM with a small number of parameters is therefore advantageous for understanding the influence of each component and fine-tuning the design. These resulting ROMs can also be extended for modeling finite-sized arrays, and the reduction in DOFs can accelerate the computation process significantly, especially in time dependent problems.

This paper is organized as follows. The general procedure of ROM development is first introduced in \cref{sec:procedure}. Then, two examples are given in \cref{sec:examples}, showing the accurately reproduced band structures. Finally, further uses of the proposed ROM approach are discussed in \cref{sec:apps}, including time domain and impact modeling of finite structures and tractable study of exceptional points and level repulsion. 
The conclusions and future outlook of this work are discussed in \cref{sec:conclusion}, where it is emphasized that the proposed approach will lead to efficient modeling and design discovery of mechanical metamaterials, with accurate construction of cellular discrete models.

\section{Formulation and Methodology}\label{sec:procedure}
\subsection{Governing equations}
The first natural step to studying a MM design is to obtain the eigenfrequency band structure.
Based on Bloch-Floquet theorem for wave propagation problems, the spatial domain is reduced to a single repeating unit cell (RUC). All fields, including the displacement field, must satisfy the Floquet periodicity:
\begin{equation}\label{eq:per}
\vec{u}(\vec{x} +  (s_1\vec{a}_1 + s_2 \vec{a}_2), t) =\vec{u}_0 (\vec{x}) \exp[\im (\omega t-\vec{k} \cdot (s_1\vec{a}_1 + s_2 \vec{a}_2))],
\end{equation}
where~$\vec{u}$ is the complex displacement vector field, the real part of which is the physical displacement vector field, $\im=\sqrt{-1}$,~$\omega=2\pi f$ is the angular frequency, $\vec{k}=[k_x,k_y]$ is the wavevector in~$x-y$ plane, $s_{1,2}$ are integers indicating different cells, and~$\vec{a}_{1,2}$ are the primitive translation vectors for a 2D unit cell. The eigenfrequency problem can be written as:
\begin{equation}\label{eq:eigf}
\left[\vec{K}(\vec{k})-\omega^2\vec{M}\right]\vec{u}_0=\vec{0},
\end{equation}
where~$\vec{K}$ and $\vec{M}$ are the stiffness and mass matrices,~$\omega^2$ is the eigenvalue, and~$\vec{u}_0$ is the eigenfunction (mode shape). The detailed development of~\cref{eq:eigf} can be found in literature \cite{Hussein2009a}. Here the matrix~$\vec{K}$ is dependent on wavevector~$\vec{k}$ since the Floquet condition is applied to the RUC. 
The Floquet condition~\cref{eq:per} and the eigenfrequency problem~\cref{eq:eigf} are the general setups used in finite element methods to find band structure and mode shapes, and have nearly identical counterparts in the ROM as well. A collection of eigen-modes can be organized in matrix $\vec{\Phi}$. Consider the~$m$-th eigen-mode solution, with frequency~$\omega_m$ and mode shape~$\vec{\Phi}_{m}$ (the $\vec{u}_0$ solution for the $m$-th mode), the time-averaged kinetic and strain energies for this mode are: 
\begin{equation}\label{eq:Tm}
\begin{split}
{\overline{T}}_m &=\frac{\omega_m}{2\pi}\int_0^{\frac{2\pi}{\omega_m}}\frac{1}{2}\Re\left[\im \omega_m e^{\im \omega_m t}\vec{\Phi}_{m}^\top\right] \Re\left[\im \omega_m e^{\im \omega_m t}\vec{M} \vec{\Phi}_{m}\right] \mathrm{d} t\\
&=\frac{1}{4}\Re[\omega_m \vec{\Phi}_{m}^\top \left(\omega_m\vec{M} \vec{\Phi}_{m}\right)^*]\\
&=\frac{1}{4}\omega_m^2 \vec{\Phi}_{m}^\dagger\vec{M} \vec{\Phi}_{m},
\end{split}
\end{equation}
\begin{equation}\label{eq:Vm}
\begin{split}
{\overline{V}}_{m}&=\frac{\omega_m}{2\pi}\int_0^{\frac{2\pi}{\omega_m}}\frac{1}{2}\Re\left[ e^{\im \omega_m t}\vec{\Phi}_{m}^\top\right] \Re\left[ e^{\im \omega_m t}\vec{K} \vec{\Phi}_{m}\right] \mathrm{d} t \ \ \ \ \ \ \ \ \\
&=\frac{1}{4}\Re[\vec{\Phi}_{m}^\top \left(\vec{K} \vec{\Phi}_{m}\right)^*]\\
&=\frac{1}{4} \vec{\Phi}_{m}^\dagger\vec{K} \vec{\Phi}_{m}
\end{split}
\end{equation}
where~$*$ and~$\dagger$ represent complex conjugate and conjugate transpose, respectively. Here \cref{eq:Tm} and \cref{eq:Vm} are valid for lossless systems for which~$\vec{K}$ and~$\vec{M}$ are Hermitian matrices.
Based on the modal orthogonality, we further have
\begin{equation}\label{eq:mt}
    \overline{\vec{T}}=\mathrm{diag}[\overline{T}_1,\dots,\overline{T}_{n_m} ]=\frac{1}{4}\vec{\Phi}^\dagger\vec{M}\vec{\Phi}\vec{\omega}^2,
\end{equation}
\begin{equation}\label{eq:mv}
    \overline{\vec{V}}=\mathrm{diag}[\overline{V}_1,\dots,\overline{V}_{n_m} ]=\frac{1}{4}\vec{\Phi}^\dagger\vec{K}\vec{\Phi}
\end{equation}
where~$\vec{\omega}=\mathrm{diag}[\omega_1,\dots,\omega_{n_m}]$ for the lowest~$n_m$ modes of interest. 
The modal energy matrices~$\overline{\vec{T}}$ and~$\overline{\vec{V}}$ are global quantities that can be evaluated in finite element solvers. 
%

\subsection{Model order reduction}
The governing equations introduced in the previous subsection are generally valid for both continuum systems and their discrete (reduced order) counterparts. 
The essential idea of the proposed ROM method, is to find the small-sized stiffness $\vec{K}^\mathrm{r}$ and mass matrices $\vec{M}^\mathrm{r}$ in such a way that the resulting global quantities~$\overline{\vec{T}}^\mathrm{r}$, $\overline{\vec{V}}^\mathrm{r}$, and~$\vec{\omega}^\mathrm{r}$ of the reduced system are preserved and directly associated with the continuum results, identified as~$\overline{\vec{T}}^\mathrm{c}$, $\overline{\vec{V}}^\mathrm{c}$, and~$\vec{\omega}^\mathrm{c}$. To achieve this, the matrix size reduction is performed by first down-sampling the continuum mode shapes~$\vec{\Phi}^\mathrm{c}$ at a set of $n_p$ primary nodal positions to obtain the sampled mode shapes $\vec{\Phi}^\mathrm{p}$, from which the ROM mode shapes~$\vec{\Phi}^\mathrm{r}$ are to be extracted.
Then, the effective ROM matrices $\vec{K}^\mathrm{r}$ and $\vec{M}^\mathrm{r}$ can be found in order to satisfy
\begin{equation}\label{eq:Tmatch}
    \overline{\vec{T}}^\mathrm{r}=\frac{1}{4}\vec{\Phi}^{\mathrm{r}\dagger} \vec{M}^\mathrm{r} \vec{\Phi}^\mathrm{r}(\vec{\omega}^\mathrm{c})^2\approx\frac{1}{4}\vec{\Phi}^{\mathrm{p}\dagger} \vec{M}^\mathrm{p}\vec{\Phi}^\mathrm{p}\vec(\vec{\omega}^\mathrm{c})^2\approx\overline{\vec{T}}^\mathrm{c}=\overline{\vec{V}}^\mathrm{c}
\end{equation}
\begin{equation}\label{eq:Vmatch}
    \overline{\vec{V}}^\mathrm{r}=\frac{1}{4}\vec{\Phi}^{\mathrm{r}\dagger} \vec{K}^\mathrm{r} \vec{\Phi}^\mathrm{r}\approx\overline{\vec{T}}^\mathrm{r}.
\end{equation}
Here the target quantities to be identified are the ROM matrices~$\vec{M}^\mathrm{r},\vec{K}^\mathrm{r}$, and other quantities~$\vec{\omega}^\mathrm{c},\vec{\Phi}^\mathrm{r},\overline{\vec{T}}^\mathrm{c},\overline{\vec{V}}^\mathrm{c}$ are obtained from continuum simulations. 
Due to the known geometric layout and domain knowledge (i.e., beam stiffness formulation), the ROM matrices are symbolically parameterized by a set of effective physical parameters that describes the structural and inertia features. In this proposed method, the effective ROM parameters are the beam stiffness parameters~$\vec{\beta}$ and the nodal inertia~$\vec{\mu}$. 
Identification of~$\vec{\beta}$ and $\vec{\mu}$ for the beam elements and nodes will complete the construction of~$\vec{K}^\mathrm{r}(\vec{\beta},\vec{k})$ and~$\vec{M}^\mathrm{r}(\vec{\mu})$.

The full procedure of ROM construction is as follows: 
\begin{enumerate}
    \item Assign a set ($n_p$) of primary nodes in the continuum unit cell, from whose displacement and rotation values the full continuum mode shapes may be approximated;
    \item Perform eigenfrequency simulations at a few~($n_k$) selected wavevectors to determine the frequency~$\vec{\omega}^\mathrm{c}$, down-sampled continuum mode shapes~$\vec{\Phi}^\mathrm{p}$, and the modal energies~$\overline{\vec{T}}^\mathrm{c}$,~$\overline{\vec{V}}^\mathrm{c}$ for the lowest~$n_m$ modes. The superscript~$^\mathrm{c}$ indicates the global quantities measured from the continuum calculations, and the superscript~$^\mathrm{p}$ denotes the down-sampled mode shapes;
    \item Construct the symbolic stiffness $\vec{K}^\mathrm{f}(\vec{\beta})$ and mass $\vec{M}^\mathrm{f}(\vec{\mu})$ matrices based on the unit cell geometry (layout, connectivity, symmetry) and positions of the~$n_p$ primary nodes as well as those of the additional~$n_d$ dependent ones (to be eliminated based on Floquet periodicity);
    \item Apply Floquet boundary condition to the symbolic matrices so that the equations of motion only involve the DOFs at the~$n_p$ primary nodes, yielding the symbolic matrices $\vec{K}^\mathrm{p} (\vec{\beta},\vec{k})$ and $\vec{M}^\mathrm{p}(\vec{\mu})$;
    \item Find the effective mass matrix~$\vec{M}^\mathrm{p}$ (i.e. parameters $\vec{\mu}$) by optimizing the kinetic energy fitness (matching the ROM results with the continuum~$\overline{\vec{T}}^\mathrm{c}$), for the lowest~$n_m$ modes, using the measured~$\vec{\omega}^\mathrm{c}$ and~$\vec{\Phi}^\mathrm{p}$; 
    \item Identify the slave DOFs (whose contribution to kinetic energy is negligible) and perform static condensation so that the number of studied DOFs is reduced (from $3n_p$ to~$n_r$). This step establishes the numerical-valued matrix~$\vec{M}^\mathrm{r}$ (a sub-matrix of $\vec{M}^\mathrm{p}$), the symbolic matrix~$\vec{K}^\mathrm{r}(\vec{\beta},\vec{k})$, and reduces the measured modes from~$\vec{\Phi}^\mathrm{p}$ to~$\vec{\Phi}^\mathrm{r}$; 
    \item Find the effective stiffness parameters~$\vec{\beta}$ by optimizing the potential energy fitness (matching the ROM modal matrix~$\overline{\vec{V}}^\mathrm{r}$ with the already determined~$\overline{\vec{T}}^\mathrm{r}$), for the lowest~$n_m$ modes, using~$\vec{\omega}^\mathrm{c}$ and~$\vec{\Phi}^\mathrm{r}$ and the established ROM mass matrix~$\vec{M}^\mathrm{r}$;  
    \item Use the established matrices~$\vec{K}^\mathrm{r}$ and~$\vec{M}^\mathrm{r}$ to compute the band structure, or adjust them for other types of problems. 
    \end{enumerate}
By matching the diagonalized modal matrices resulting from the symbolic discrete model and the FEM model, this process aims to construct a discretized lower order system that, at the selected wavevector~$\vec{k}$ points, inherits the continuum eigenvalues~$\vec{\omega}^\mathrm{c}$ and the associated mode shapes~$\vec{\Phi}^\mathrm{r}$ down-sampled from the continuum system. 
With the ROM matrices $\vec{K}^\mathrm{r}(\vec{\beta},\vec{k})$ and~$\vec{M}^\mathrm{r}(\vec{\mu})$ symbolically parameterized instead of numerically found through the pseudo-inversion of \crefrange{eq:Tmatch}{eq:Vmatch}, the known structural knowledge is retained in the model. This allows for further analysis and optimization of the structural components in design efforts. 
Since the number of these unknown parameters in~$\vec{\beta}$ and~$\vec{\mu}$ are limited, one does not need to optimize the matching of \crefrange{eq:Tmatch}{eq:Vmatch} for every wavevector~$\vec{k}$. Instead, the eigen-information from only a small number of wavevectors will be sufficient to determine the unknown ROM parameters and the number of needed simulations is small.
In addition, the extracted stiffness ($\vec{\beta}$) and inertia~($\vec{\mu}$) values for the discrete system are properly scaled to represent effective physical quantities due to the matched modal energies. Since the effective parameters~$\vec{\beta}$ and~$\vec{\mu}$ are of high physical fidelity and independent of wavevector~$\vec{k}$, one can easily compute the eigen-results at any arbitrary wavevector with the ROM matrices. Furthermore, the ROM can be easily extended for other types of computations, such as frequency or time domain problems, for which finite-sized arrays are modeled, and wavevector~$\vec{k}$ is not an explicit parameter.
In summary, such a model order reduction approach serves as both a parameter retrieval method that characterizes the continuum model as a discrete one, as well as a fast tool that accelerates the computation of eigen- or other dynamic problems.

In the next section of this paper, the detailed ROM construction steps are demonstrated through examples.  

\section{Parameter Extraction Procedure}\label{sec:examples}
\subsection{Node assignment and information collection}
\begin{figure}[!ht]
	\begin{subfigure}[b]{0.5\linewidth}
		\centering\includegraphics[height=110pt]{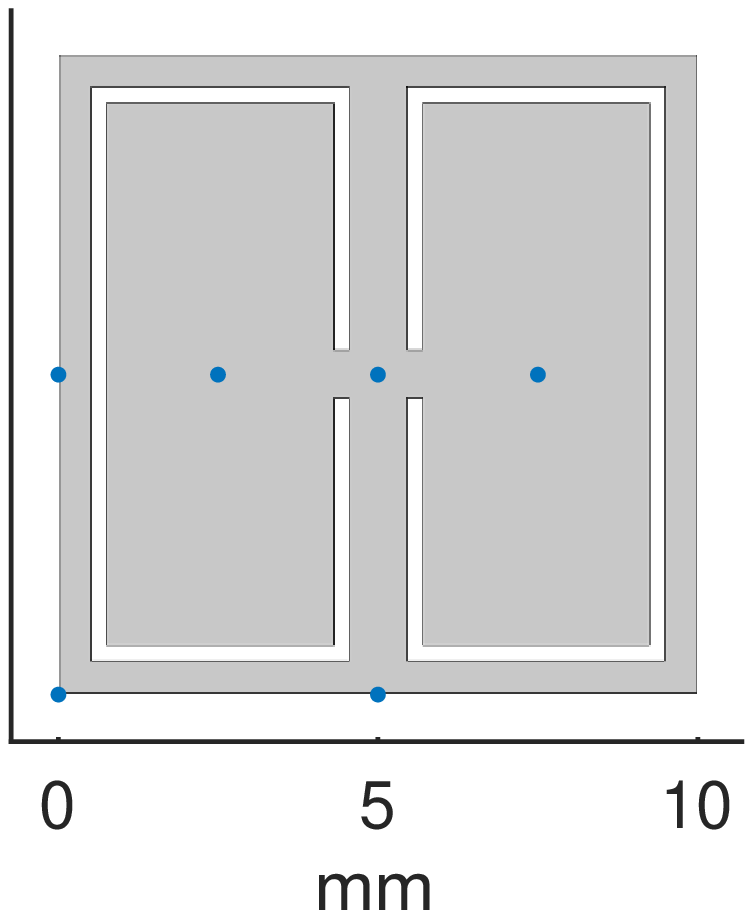}
		\caption{Square MM \label{fig:sh_geom}}
	\end{subfigure}%
	\begin{subfigure}[b]{0.5\linewidth}
		\centering\includegraphics[height=110pt]{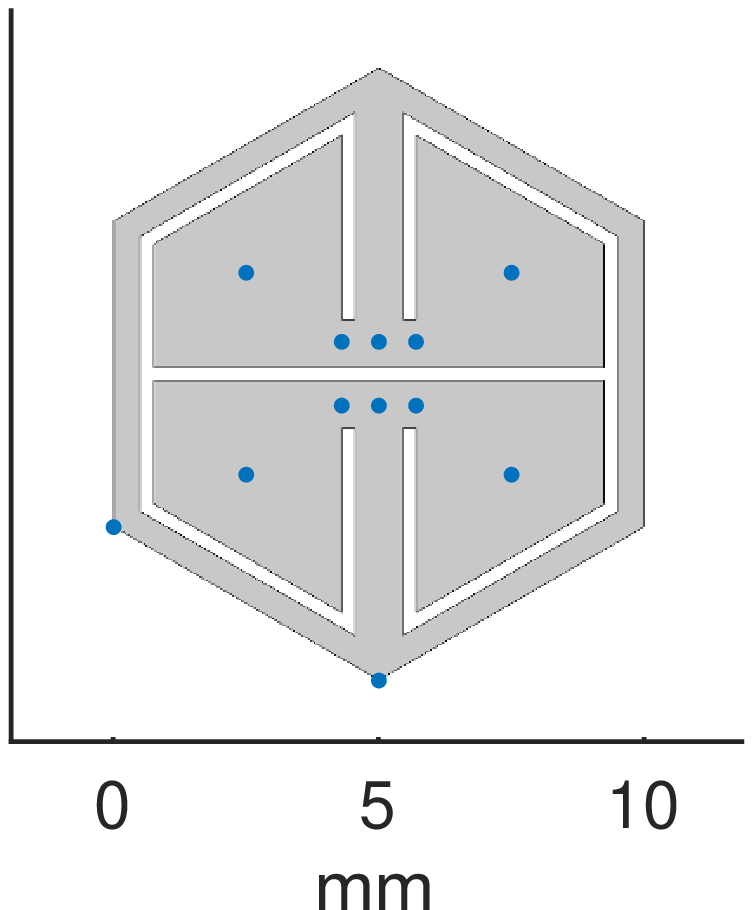}
		\caption{Hexagonal MM\label{fig:h0_geom}}
	\end{subfigure}
	
	\caption{Unit cell geometry of the selected examples. The blue dots denote the selected~$n_p$ nodes where the mode shapes~$\vec{\Phi}^\mathrm{p}$ are sampled from FEM solutions at selected wavevectors. 
	\label{fig:geom}} 
\end{figure}

To demonstrate the ROM parameterization process, two MM unit cells are selected. The square unit cell \Cref{fig:sh_geom} features an H-shaped resonator mass, while the hexagonal cell \cref{fig:h0_geom} has two split resonators. The two RUCs are modeled in FEM using the same material (typical alumina), with Young's modulus~$E=\SI{300}{GPa}$, Poisson's ratio~$\nu=0.22$, and density~$\rho=\SI{3900}{kg/m^3}$. Both designs have the lattice constant~$a=\SI{10}{mm}$. One should first assign a set of $n_p$ nodes in the continuum model, where the deformation will be sampled and used for mode shape projections. These sampling points should, in principle, be located at the mass centers of structural components, or at the intersections between beam elements.
Further detailed analysis and principles of DOF selection are given in literature \cite{Kevin2015,Qu2004}.
The chosen sampling nodes in the two examples are denoted by the blue dots in \cref{fig:geom}. Notice that there is no node assigned at the top or right edge of the cell frames, due to the known Floquet periodicity~\cref{eq:per} for infinite arrays. The mode shapes~$\vec{\Phi}^\mathrm{p}$ will be represented by the deformation vector~$\vec{u}^\mathrm{p}$ containing the displacement~$u_x,u_y$ in~$x-y$ plane and the rotation~$\theta_z$ at the chosen nodes. The rotational component is derived from the curl of the continuum displacement field~$\theta_z=(\frac{\partial u_y}{\partial x}-\frac{\partial u_x}{\partial y})/2$.

Then, one performs finite element simulations at a few selected~$\vec{k}$ points in the irreducible Brillouin zone (IBZ). Preferably, these~$\vec{k}$ points should be far away from each other. In the shown examples here, we select~$n_k=4$ points at~$\vec{k}=[0,0],[\pi/a,0],[0,\pi/a],[\pi/a,\pi/a]$. At each wavevector point, one collects the eigenfrequency results for the lowest~$n_m$ modes. The number~$n_m$ is determined such that: (1) the~$n_m$-th frequency~$f_{n_m}$ covers the frequency range of interest and (2) the locations associated with dominant deformations for the lowest~$n_m$ modes are included in the pre-selected nodes. 
We chose~$n_m=6$ for the square cell and~$n_m=8$ for the hexagonal one. In this step, the collected information includes the diagonal frequency matrix~$\vec{\omega}^\mathrm{c}\in\mathbb{R}^{n_m\times n_m}$, the diagonal kinetic energy matrix~$\overline{\vec{T}}^\mathrm{c}\in\mathbb{R}^{n_m\times n_m}$, the diagonal potential energy matrix~$\overline{\vec{V}}^\mathrm{c}\in\mathbb{R}^{n_m\times n_m}$, and the mode shape matrix~$\vec{\Phi}^\mathrm{p}\in\mathbb{C}^{3n_p\times n_m}$. 

\subsection{Matrix construction}
\begin{figure}[!ht]
	\begin{subfigure}[b]{0.5\linewidth}
		\centering\includegraphics[height=80pt]{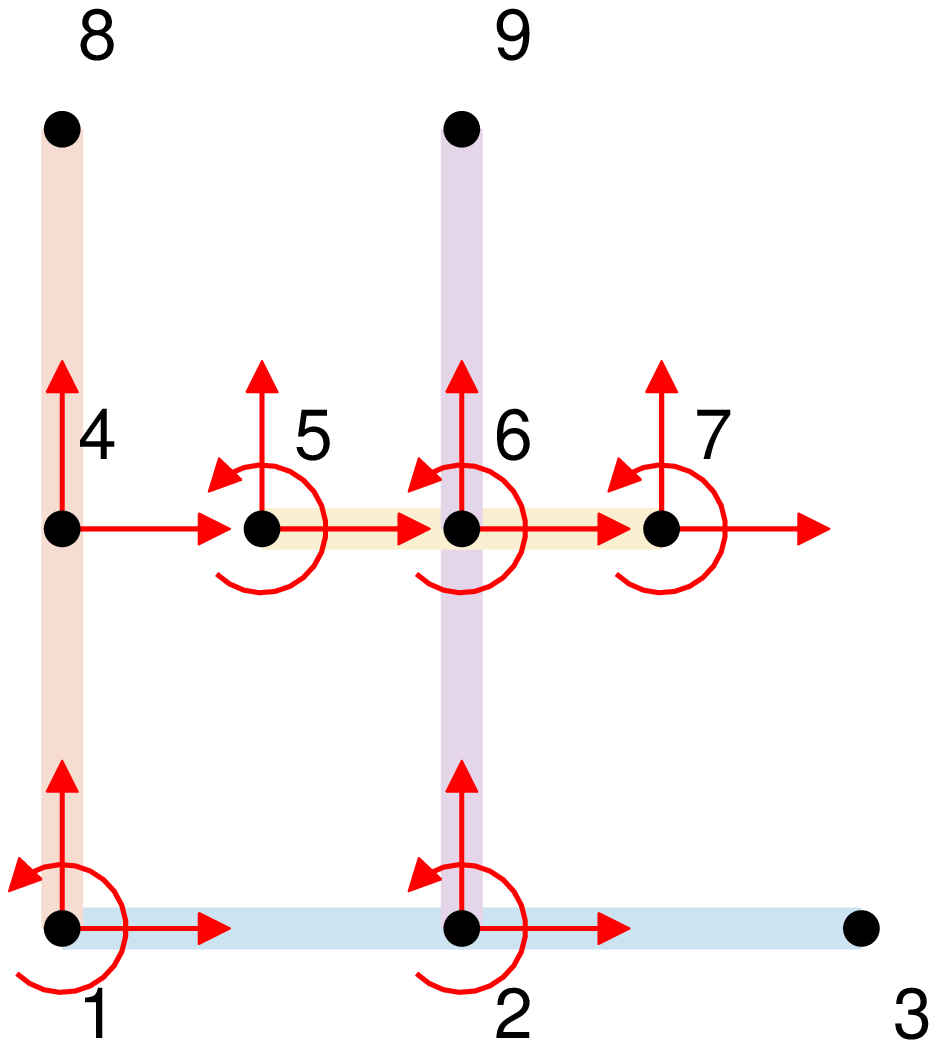}
		\caption{\label{fig:sh_ruc_dofs}}
	\end{subfigure}%
	\begin{subfigure}[b]{0.5\linewidth}
		\centering\includegraphics[height=80pt]{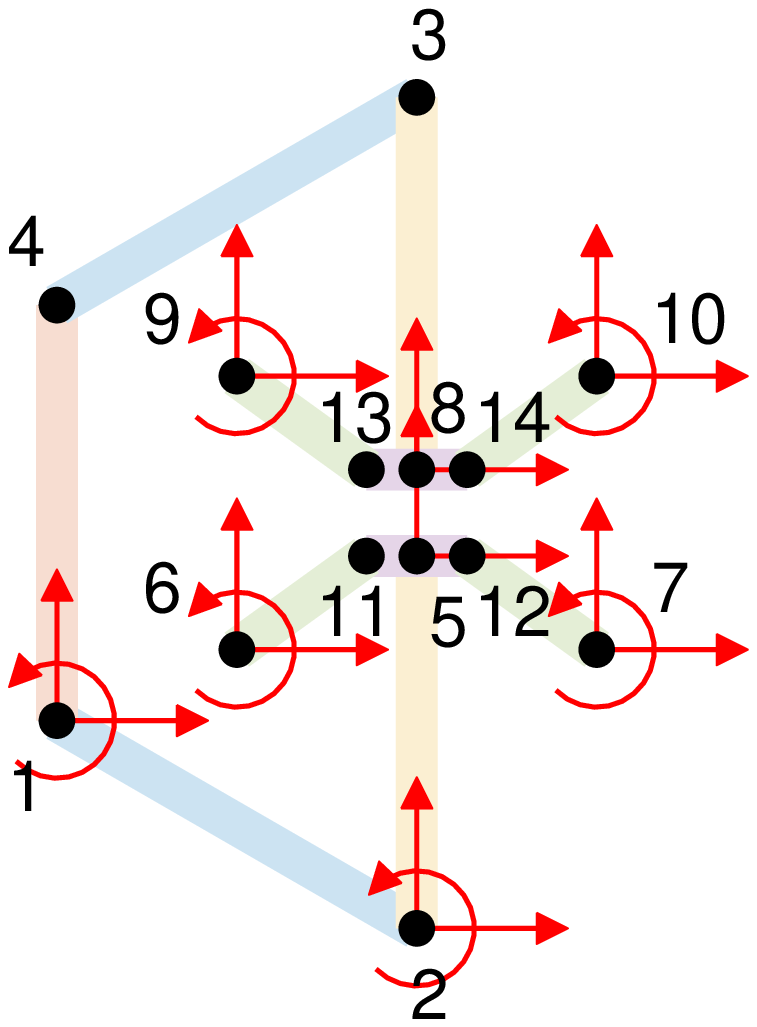}
		\caption{\label{fig:h0_ruc_dofs}}
	\end{subfigure}
	\caption{ Beam assemblies as the reduced order unit cells for the (\subref{fig:sh_ruc_dofs}) square and (\subref{fig:h0_ruc_dofs}) hexagonal MMs. The red arrows denote the reduced inertia DOFs.  \label{fig:rom_ruc_dof}} 
\end{figure}

After the FEM data collection, the ROM matrices can be constructed symbolically based on the selected nodes, including the~$n_p$ primary ones and the~$n_d$ dependent ones, based on the known connectivity and beam element stiffness formulation (see Appendix \cref{eq:beamK}). The dependent nodes are those whose displacements are determined based on Floquet periodicity, yet are connected by a structural element to one of the primary nodes. The graphic representations of the ROM unit cells are shown in \cref{fig:rom_ruc_dof}. 
For the square unit cell in~\cref{fig:sh_ruc_dofs}, the dependent DOFs are~$\vec{u}^\mathrm{d}=[\vec{u}^{(3)},\vec{u}^{(8)},\vec{u}^{(9)}]^\top$. For the hexagonal unit cell in~\cref{fig:h0_ruc_dofs}, the dependent DOFs are~$\vec{u}^\mathrm{d}=[\vec{u}^{(3)},\vec{u}^{(4)}]^\top$.
Notice that some edge elements (between dependent nodes) are not included in order to eliminate redundancy in the periodically generated array. For example, nodes 8 and 9 are not directly connected in \cref{fig:sh_ruc_dofs}. 
Nevertheless, the structures shown in \cref{fig:rom_ruc_dof} are primitive unit cells whose 2D repetitions will produce the infinitely periodic system perfectly. 
Each node (denoted by a black dot) has three inertia parameters~$m_x,m_y=m_x$ (mass) and~$I_z$ (rotational inertia). The force-balance relation between two connected nodes is approximated based on beam analysis, introduced in \cref{appBeam}.
In this formulation, the beam element is allowed to have an asymmetric layout, nevertheless only four independent stiffness parameters~$\beta_{1,2,5,7}$ (diagonal components of the stiffness matrix) are to be determined in order to construct the local stiffness matrix as the other components are statically determined. 
Such a form is not only compatible with standard beam elements (Euler-Bernoulli, Timoshenko), but also suitable for any generalized 1D structural component with two end nodes. 
To obtain the global stiffness matrix, each force-balance relation is first converted into the global coordinate system; See~\cref{appBeam}. Based on the equilibrium of the overall structure, the global stiffness matrix is obtained by summing all the loads arising from the adjacent elements for each node \cite{ferreira2008matlab}. The static balance equations then read
\begin{equation}
    \vec{K}^\mathrm{f}\vec{u}^\mathrm{f}=\vec{F},
\end{equation}
where
\begin{equation}
    \vec{F}=\begin{bmatrix}
    u^{(1)}_x& u^{(1)}_y& \theta^{(1)}_z&\cdots& \theta^{(n_p+n_d)}_z
    \end{bmatrix}^\top
\end{equation}
is the nodal loading and superscript is the node index. The symbolic matrix~$\vec{K}^\mathrm{f}$ gives the constitutive description of the full unanchored structure, with free boundary conditions. The associated mass matrix is a diagonal matrix~$\vec{M}^\mathrm{f}=\mathrm{diag}[\vec{\mu}]=\mathrm{diag}[m^{(1)}_x,\dots,I^{(n_p+n_d)}_z]$. Then, the unit cell structure is effectively parameterized by the unknown stiffnesses~$\vec{\beta}$ and inertia values $\vec{\mu}$.

To apply the Bloch-Floquet periodicity condition, one can first write the full set of DOFs~$\vec{u}^\mathrm{f}$ in terms of the DOFs at the primary nodes~$\vec{u}^\mathrm{p}$: 
\begin{equation}
    \vec{u}^\mathrm{f}=\vec{P}(\vec{k})\vec{u}^\mathrm{p},
\end{equation}
where~$\vec{P}(\vec{k})$ is a rectangular transformations matrix containing phase differences between the dependent and primary DOFs, determined by the nodal positions and the wavevector. A detailed discussion on applying the periodicity can be found in \cite{Krattiger2018a}. Then
\begin{equation}\label{eq:eom_periodic}
    \vec{K}^\mathrm{p} (\vec{k}) \vec{u}^\mathrm{p}-\vec{\omega}^2 \vec{M}^\mathrm{p}\vec{u}^\mathrm{p}=\vec{0},
\end{equation}
where~$\vec{K}^\mathrm{p}=\vec{P}^\dagger \vec{K}^\mathrm{f}\vec{P}$,~$\vec{M}^\mathrm{p}=\vec{P}^\dagger \vec{M}^\mathrm{f}\vec{P}$ are the matrices for the primitive cell, and~$\dagger$ is Hermitian transpose. At this stage, the equations of motion \cref{eq:eom_periodic} effectively describe the dynamics of the discretized primitive unit cells, and the involved nodes are identical to the ones marked in \cref{fig:geom}.

Taking advantage of the geometrical symmetries in the continuum model, the number of unknown parameters in the ROM property matrices can be reduced. For example, beam 5-6 and beam 7-6 are symmetric with respect to node 6 in~\cref{fig:sh_ruc_dofs}. Then, node 5 will have the same mass and rotational inertia as node 7. The two beams will share the same local stiffness matrix as well (in the un-rotated local coordinate system). Under this idealized description, the structural symmetries lead to degeneracy in the eigenvalues. However, any physical or numerical realization of such systems will have the tendency to become non-degenerate due to any small asymmetry. The benefits of ROM for understanding the physics of modal degeneracy will be discussed in \cref{sec:lr} and \cref{app:geometric_phase}. 

\subsection{Inertia quantification and DOF reduction}

With the symbolic ROM matrices developed, the next step is to find the effective mass and inertia values of the selected nodes. Consider the kinetic energy formulation at the~$(j)$-th $\vec{k}$ point, the ROM values are expected to be identical to the continuum ones:
\begin{equation}\label{eq:mt_e}
\frac{1}{4}(\vec{\omega}^\mathrm{c}_{(j)})^2\vec{\Phi}^{\mathrm{p}\dagger}_{(j)}\vec{M}^\mathrm{p}(\vec{\mu})\vec{\Phi}^\mathrm{p}_{(j)}
 =  \overline{\vec{T}}^\mathrm{c}_{(j)}\in\mathbb{R}^{n_m\times n_m}.
\end{equation}
As is indicated by the rhs of this equation, this is an $n_m \times n_m$ matrix equation, for each $(j)$ value associated with a point in IBZ. Its purpose is to find the  matrix~$\vec{M}^\mathrm{p}(\vec{\mu})$, assumed diagonalizable by the down-sampled eigenvectors~$\vec{\Phi}^\mathrm{p}_{(j)}$, based on the known $n_m \times n_m$ diagonal eigenfrequency matrix~$\vec{\omega}^\mathrm{c}_{(j)}$ and calculated continuum kinetic energy~$\overline{\vec{T}}^\mathrm{c}_{(j)}$. 
As this is clearly mathematically over-determined, an error vector can be defined as
\begin{equation}\label{eq:masserror}
    \vec{e}_{(j)} (\vec{\mu})=\mathrm{UT}\left[\frac{1}{4}(\vec{\omega}^\mathrm{c}_{(j)})^2\vec{\Phi}^{\mathrm{p}\dagger}_{(j)}\vec{M}^\mathrm{p}(\vec{\mu})\vec{\Phi}^{\mathrm{p}}_{(j)}
 -  \overline{\vec{T}}^\mathrm{c}_{(j)}\right],
\end{equation}
where~$\mathrm{UT}[\cdot]$ denotes the vector containing all the upper triangular entries of a matrix. 
Combining the results at all the~$n_k$ wavevectors, the error vector is then $\vec{E}=[\vec{e}_{(1)}\dots\vec{e}_ {(n_k)}]$. 
This problem can be stated as a constrained linear least-squares problem
\begin{equation}
\begin{aligned}
\min_{\vec{\mu}} \quad & ||\vec{E}(\vec{\mu})||_2^2\\
\textrm{s.t.} \quad & \vec{\mu} \geq 0 \\
\end{aligned}
\end{equation}
where $||\cdot||_2$ indicates the~$L_2$ norm, and the condition on $\vec{\mu}$ is understood to apply to each component independently. In practice, to guarantee the equal participation of each mode and each wavevector, all the mode shapes should be pre-normalized so that their kinetic energies are equal.
Nevertheless, the off-diagonal components in the modal matrix~$\overline{\vec{T}}^\mathrm{c}$ will remain zero. 
The optimization problem is solved using the \textit{lsqlin} function in MATLAB\textsuperscript{\textregistered}. 
For both models, the optimization leads to good convergence with the error ${\vec{\mu}}$ less than~$3\%$.
By using the simulation results at more than one ~$\vec{k}$ points (in this case,~$n_k=4$), the real and imaginary parts of the upper triangular components together lead to ($n_k n_m (n_m+1)$) real equations for finding the unknown inertia parameters.
Furthermore, the lowest two modes at~$\vec{k}=\Gamma=[0,0]$ are rigid body modes with zero eigenfrequencies. The inclusion of the~$\Gamma$ point in this process will automatically guarantee that the solved solution satisfies mass conservation. 

\begin{figure}[!ht]
	\begin{subfigure}[b]{1\linewidth}
		\centering\includegraphics[height=110pt]{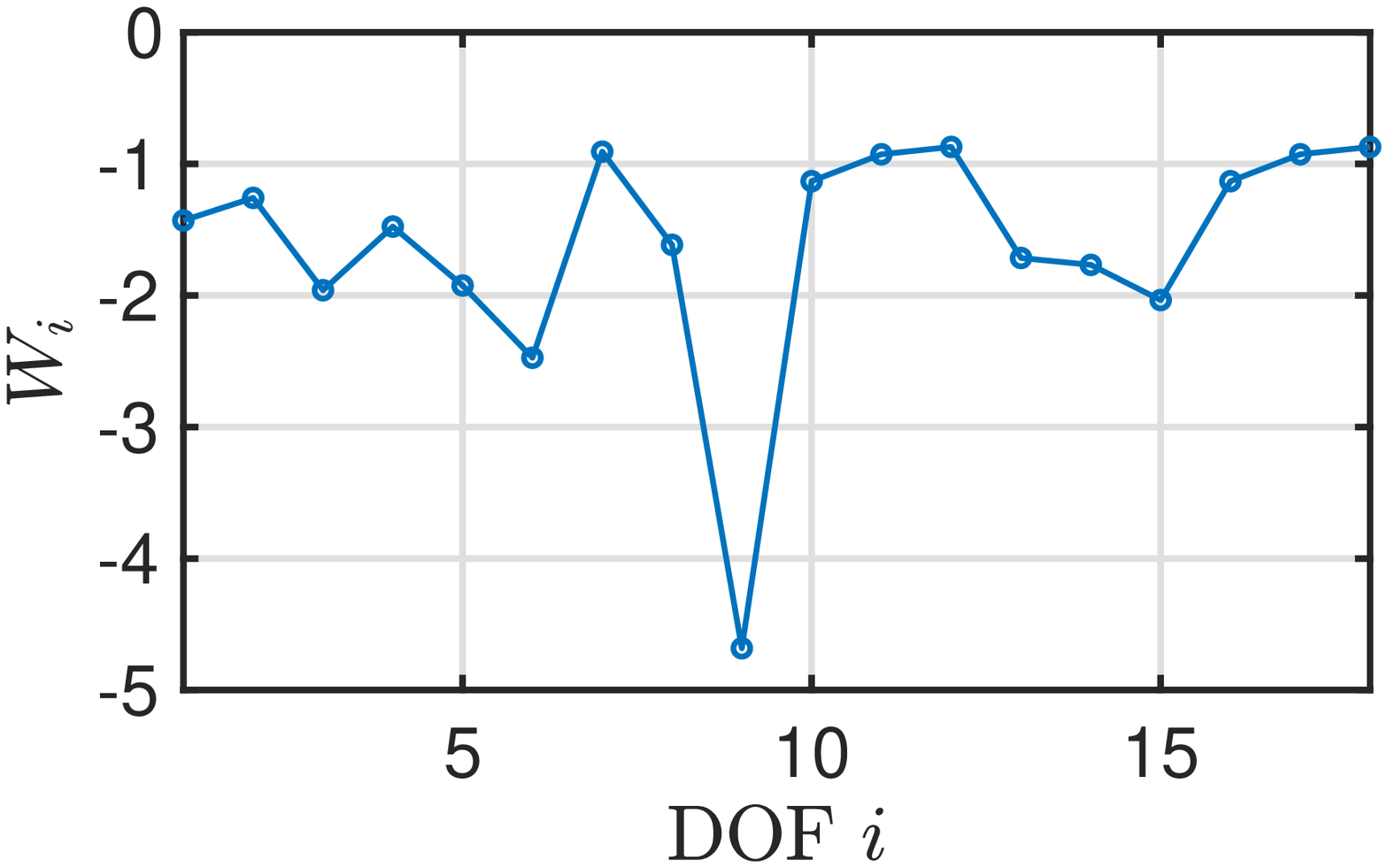}
		\caption{\label{fig:sh_dofwei}}
	\end{subfigure}\\
	\begin{subfigure}[b]{1\linewidth}
		\centering\includegraphics[height=110pt]{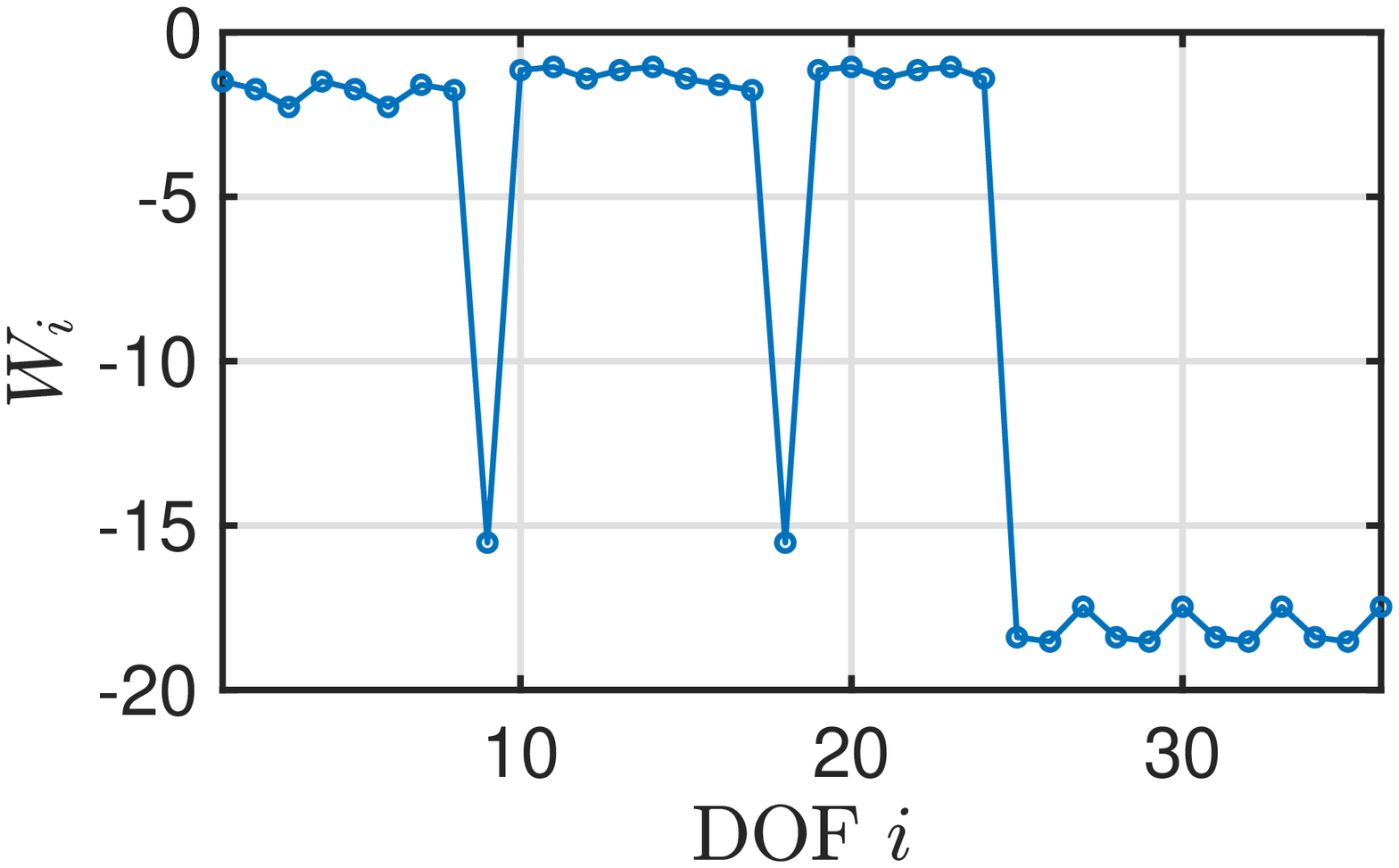}
		\caption{\label{fig:h0_dot_wei}}
	\end{subfigure}
	\caption{Kinetic importance weights of the DOFs (in log scale) for the (\subref{fig:sh_dofwei}) square and (\subref{fig:h0_dot_wei}) hexagonal MMs. \label{fig:dof_weights}} 
\end{figure}

Although three DOFs ($u_x,u_y,\theta_z$) are sampled at each node, not all of them have the same importance. Certain DOFs may only affect the mode shapes only when the frequency is high enough, and certain DOFs may be associated with negligible inertia. With the mass matrix found, the mode shapes~$\vec{\Phi}^\mathrm{p}$ can be re-normalized so that
\begin{equation}\label{eq:Tnorm}
   \frac{1}{4} (\omega^\mathrm{c}_{m(j)})^2  \vec{\Phi}^{\mathrm{p}\dagger}_{m(j)}\vec{M}^\mathrm{p}\vec{\Phi}^\mathrm{p}_{m(j)}=1 \  \qquad \mathrm{for\ any\ }m,(j)
\end{equation}
where subscript~$_{m(j)}$ indicates the~$m$-th mode eigenvalue at~$(j)$-th wavevector location.
Then, the kinetic importance weight of the~$i$-th DOF is evaluated as the averaged kinetic energy in the log scale:
\begin{equation}
    W_i=\log_{10}\dfrac{\sum_{m=1}^{n_m}\sum_{j=1}^{n_k} \frac{1}{4} (\omega^\mathrm{c}_{m(j)})^2  
    \Phi^{\mathrm{p}*}_{im(j)}
    M^{\mathrm{p}}_{ii}
      \Phi^{\mathrm{p}}_{im(j)}}
  {n_m n_k}.
\end{equation}
\Cref{fig:dof_weights} shows the relative weights of all the DOFs for the two MM examples. It is apparent that certain DOFs with weight~$\leq-4$ should be eliminated and be regarded as ``slave'' DOFs, otherwise, and particularly for time domain dynamics analysis they will cause numerial challenges. For example, the ninth DOF of the square MM (the rotation at node 4) has negligible weight and is not active in the considered frequency range. Only the active ``master'' DOFs will be kept in the ROM formulations, and they are indicated by the red arrows in~\cref{fig:rom_ruc_dof}. 
Deleting the slave DOFs from~$\vec{\Phi}^\mathrm{p}$ leads to the reduced mode shapes~$\vec{\Phi}^\mathrm{r}$, which is a subset of continuum data and is expected to be the eigenvectors of the ROM matrices. 
The removal of these slave DOFs follows the standard static condensation \cite{GUYAN1965}, as the associated inertia values are effectively zero. One can re-arrange the stiffness matrix as
\begin{equation}\label{eq:KP}
\vec{K}^\mathrm{p}=\begin{bmatrix}
\vec{K}_{\mathrm{mm}} & \vec{K}_{\mathrm{ms}} \\
\vec{K}_{\mathrm{sm}} & \vec{K}_{\mathrm{ss}}
\end{bmatrix}.
\end{equation}
Here the subscripts denote ``m"aster (not to be confused with index $m$ used earlier to indicate modes) and ``s"lave index arrays.
The columns and rows related to slave DOFs in the mass matrix~$\vec{M}^\mathrm{p}$ are deleted, and it leads to the reduced mass matrix~$\vec{M}^\mathrm{r}$. The reduced (still symbolic) stiffness matrix is obtained by
\begin{equation}\label{eq:guyan}
    \vec{K}^\mathrm{r}=\vec{K}_{\mathrm{mm}}-\vec{K}_{\mathrm{ms}}\vec{K}_{\mathrm{ss}}^{-1}\vec{K}_{\mathrm{sm}}.
\end{equation}
In practice, the inversion of symbolic sub-matrix~$\vec{K}_{\mathrm{ss}}$ is computationally challenging. However, this step can be equivalently implemented using Gaussian elimination.

\subsection{Stiffness parameter extraction}
\begin{figure}[!ht]
	\begin{subfigure}[b]{1\linewidth}
		\centering\includegraphics[height=110pt]{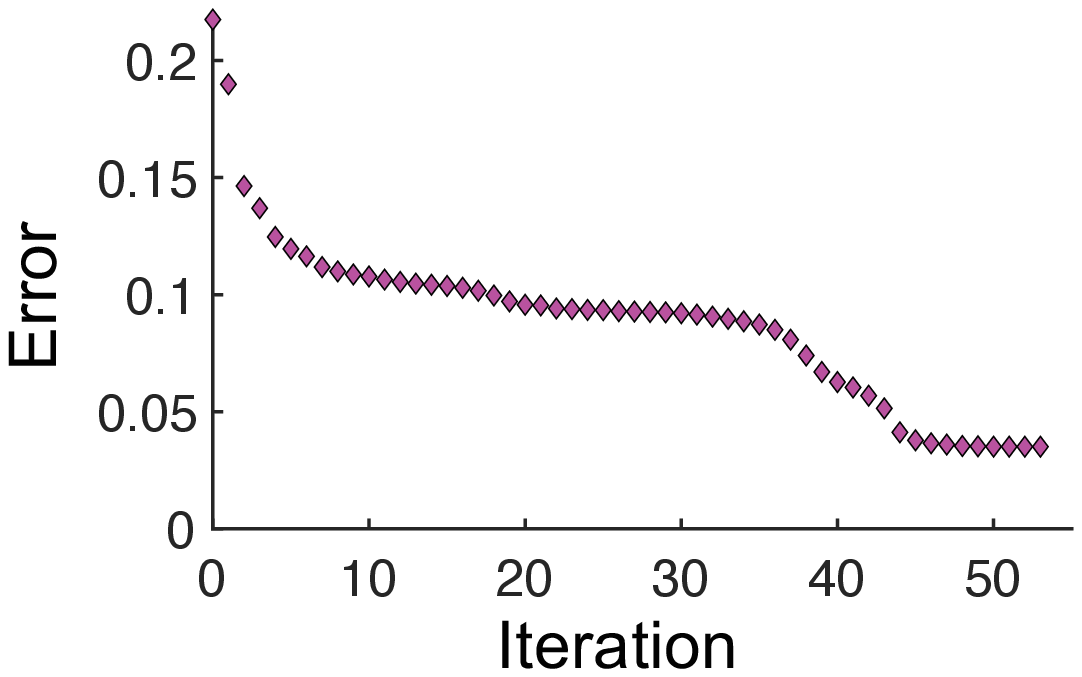}
		\caption{\label{fig:sh_beta}}
	\end{subfigure}\\
	\begin{subfigure}[b]{1\linewidth}
		\centering\includegraphics[height=110pt]{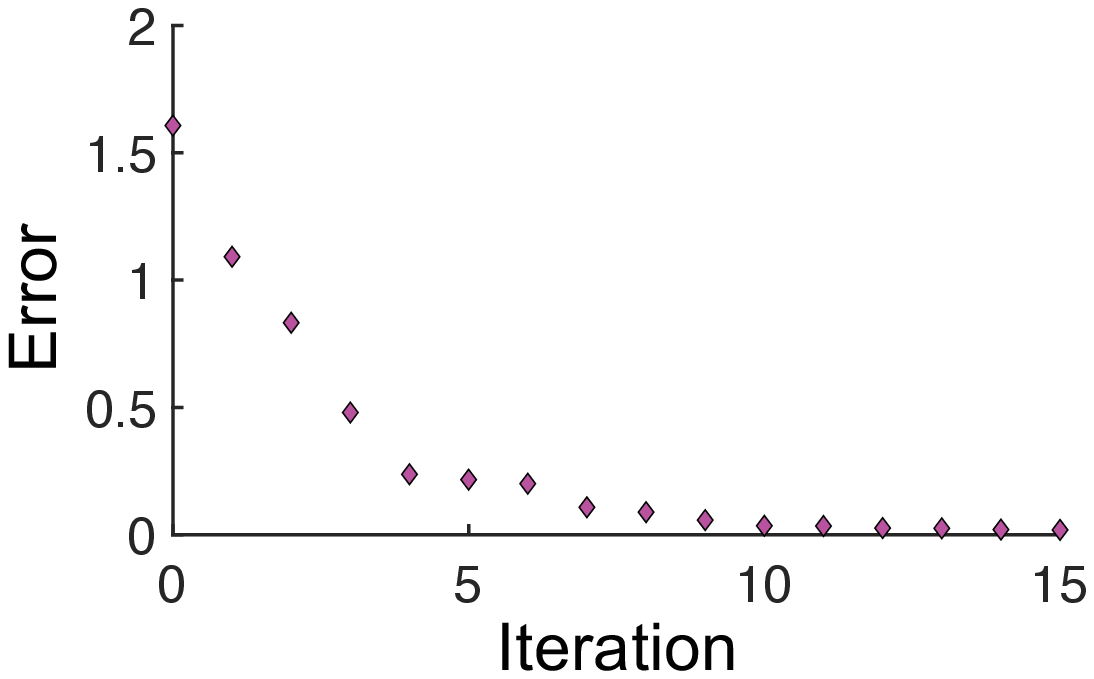}
		\caption{\label{fig:h0_beta}}
	\end{subfigure}
	\caption{Potential energy fitness optimization convergence plots for (\subref{fig:sh_beta})  square and (\subref{fig:h0_beta}) hexagonal  unit cells. \label{fig:beta}} 
\end{figure}

To ensure that the continuum eigenfrequencies and mode shapes can be accurately reproduced by the ROM, the modal potential energy must be equal to the modal kinetic energy. 
Such equality can be proved by left multiplying the mode shape in \cref{eq:eigf}.
The ideal set of stiffness parameters~$\vec{\beta}$ can therefore be found by optimizing the potential energy fitness. The error in energy at the~$(j)$-th wavevector location is defined as 
\begin{equation}\label{eq:betaerror}
    \vec{e}_ {(j)}(\vec{\beta})=\mathrm{UT}\left[\frac{1}{4}\vec{\Phi}^{\mathrm{r}\dagger}_{(j)}\vec{K}^\mathrm{r}(\vec{\beta})\vec{\Phi}^\mathrm{r}_{(j)} -\frac{1}{4}\vec{\Phi}^{\mathrm{r}\dagger}_{(j)}\vec{M}^\mathrm{r}\vec{\Phi}^\mathrm{r}_{(j)}(\vec{\omega}^\mathrm{c}_{(j)})^2  \right],
\end{equation}
where the superscript~$^r$ denotes quantities associated with the reduced set of master DOFs. The error vector for all~$n_k$ wavevector locations is then~$\vec{E}=[\vec{e}_{(1)}\dots\vec{e}_{(n_k)}]$. 
The optimization problem is then formulated as:
\begin{equation}\label{eq:Vopt}
\begin{aligned}
\min_{\vec{\beta}} \quad & {||\vec{E}(\vec{\beta})||_2}\\
\textrm{s.t.} \quad & \vec{\beta}>\vec{0},   \\
\end{aligned}
\end{equation}
where the constraint on $\vec{\beta}$ is understood as positivity for every single $\beta$ parameter in the structure; See \cref{appBeam} for detail. 
This process also ensures the diagonality of the potential energy in ROM formulation.
The mode shapes are normalized based on~\cref{eq:Tnorm} to ensure equal weights in all the modes. Notice that due to the static condensation process \cref{eq:guyan}, the stiffness matrix~$\vec{K}^\mathrm{r}$ and the error vector~$\vec{E}$ are no longer linear functions of the stiffness~$\vec{\beta}$. Furthermore, the stiffness parameters need to be rescaled properly due to numerical considerations. For example, the rotational stiffnesses have different units than axial or translational stiffnesses. Therefore, it is beneficial to express the stiffness as
\begin{equation}
    \vec{\beta}=\vec{\alpha}\circ\hat{\vec{\beta}}
\end{equation}
where~$\circ$ is the element-wise multiplication,~$\vec{\alpha}$ is a dimensionless stiffness ratio vector, and the vector~$\hat{\vec{\beta}}$ contains the estimated stiffness values based on the beam geometry (length, height), which can be derived using the standard formulas of Timoshenko beam theory. Then \cref{eq:Vopt} can be re-written as
\begin{equation}
\begin{aligned}
\min_{\vec{\alpha}} \quad & {||\vec{E}(\vec{\alpha})||_2}\\
\textrm{s.t.} \quad & \vec{\alpha}>\vec{0} .   \\
\end{aligned}
\end{equation}
Such a problem can be initialized from~$\vec{\alpha}=\vec{1}$ and is solvable using the \textit{fmincon} function in MATLAB\textsuperscript{\textregistered}. Furthermore, effects of deviation from these initial estimates are of the similar order of magnitude, which is numerically preferred. \Cref{fig:beta} shows the error (cost) convergence for the two examples considered here. For the square MM~\cref{fig:sh_beta}, it takes more iterations to reach the minimum. However, both cases present decently low errors in the final iterations.

\subsection{Verification and discussion}

With the effective stiffness~$\vec{\beta}$ and inertia~$\vec{\mu}$ parameters determined, the ROM procedure is completed. The optimized modal energy fitness ensures the fidelity of the ROM. It is observed that the optimized matching of modal relation leads to the accurate reproduction of the eigenfrequencies as well as the mode shapes at the pre-calculated~$n_k$ wavevector locations. Beyond these locations, the eigen-analysis results are also well extrapolated because of the symbolic implementation of the structural stiffness and Bloch-Floquet periodicity.
One can solve for the eigenfrequency and mode shapes through the analytical formulation~\cref{eq:eigf} for any given wavevector~$\vec{k}$. Such computation will be extremely fast due to the compactness of the matrices. 
\begin{figure}[!ht]
	\begin{subfigure}[b]{1\linewidth}
		\centering\includegraphics[height=210pt]{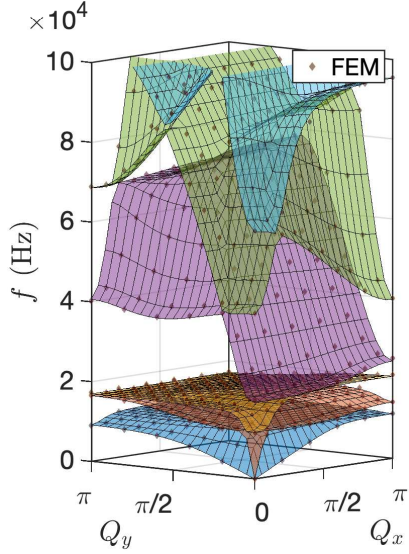}
		\caption{Square MM\label{fig:sh_band}}
	\end{subfigure}\\
	\begin{subfigure}[b]{1\linewidth}
		\centering\includegraphics[height=210pt]{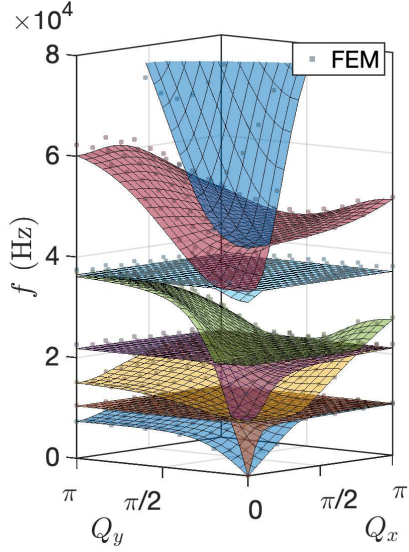}
		\caption{Hexagonal MM\label{fig:h0_band}}
	\end{subfigure}
	\caption{Band structure comparison.  \label{fig:band}} 
\end{figure}
\Cref{fig:band} shows the eigenfrequency band structures for the two studied examples, plotted in the dimensionless wavenumber space~$Q_{x,y}=k_{x,y}a$, where~$a=\SI{10}{mm}$ is the lattice constant. The colored surfaces are generated based on ROM, while the dots represent FEM results. It can be seen that the ROM provides close approximations of the band structures. Notice that the ROM construction only requires the simulations at four different wavevector locations $[Q_x,Q_y]=[0,0],[0,\pi],[\pi,0]$ and $[\pi,\pi]$. The number and locations of these input simulations are, however, not fixed to the given ones. 

It should be noted that the minimizing the matching error~$\vec{e}_{(j)}(\vec{\beta})\rightarrow\vec{0}$ in \cref{eq:betaerror} is a necessary yet insufficient requirement for the ROM system~$\vec{K}^\mathrm{r}(\vec{\beta}),~\vec{M}^\mathrm{r}$ to produce the exact eigen-solutions~$\vec{\Phi}^\mathrm{r}_{(j)},~\vec{\omega}^\mathrm{c}_{(j)}$ obtained from FEM simulations. The left multiplication of~$\vec{\Phi}^\mathrm{r}_{(j)}$ in \cref{{eq:betaerror}} reduces the number of equations since the mode shape matrix is rectangular. However, the number of unknowns in~$\vec{\beta}$ is limited, and \cref{eq:betaerror} is collected for multiple wavenumber points. Therefore, such an optimization scheme creates an over-determined problem for seeking the limited set of ROM parameters that are representative of the unit cell properties. With known eigen-states, the ROM produces the same modal energy matrices as the higher order FEM system. Then, it is observed that the resulting ROM leads to eigenvectors that closely agree with the FEM results. An alternative way to identify~$\vec{\beta}$ is to create a multi-objective optimization problem in which one also optimizes the ROM mode shape accuracy while minimizing the modal energy error given by \cref{eq:betaerror}. In practice and in the examples here, the secondary optimization objective of maintaining mode shape accuracy is omitted and only used as a sanity check, leading to significant computational cost savings but insignificant or no loss to accuracy. The latter outcome is understood to be a consequence of the symbolic development of dynamic matrices based on the internal cell topology (beam connectivity).

This approach advances the well-established model order reduction methods such as SEREP \cite{Avitable1989} in attacking MM problems in the sense that (1) the proposed ROM maintains the eigen-solution accuracy for any wavevector and (2) it provides analytical and parameterized matrices instead of numerical ones. It expands such methods by including the propagating nature of waves instead of modal response of finite structures. In MM systems, the micro-structural features play vital roles in the dynamic properties. A small variation in the geometry could lead to a drastic change in the overall response. Therefore, an analytical model with parameterized structural elements is particularly advantageous for understanding the influence of each component and fine-tuning the design. Several applications of the method are discussed in the next section.

\section{Applications}\label{sec:apps}

The proposed ROM approach has a wide application spectrum, as the matrices are parameterized by the physical properties (structural stiffness and inertia) and the modeled DOFs are physical deformations instead of generalized coordinates. Therefore the developed ROMs preserve the necessary physical ingredients for further analysis. The dependence of ROM parameters on the geometric dimensions and material properties may be curve-fitted for design purposes with continuous functions, see \cite{Morris2022}. The fidelity of these models is inherently guaranteed by the optimized energy relations and the accurate production of band structures and mode shapes. While the ROM is capable of generating the band structure accurately and efficiently, the band computation is not the ultimate goal of the ROM, rather it is the basis and a starting point.

One immediate application is optimizing unit cell designs for desired eigenfrequencies. The ROM characterizes the continuum unit cell with a finite number of stiffness and inertia parameters and provides the analytical formulation of the eigenfrequency bands. It is then intuitive to tune the structural parameters and associated geometry for desired eigenfrequency performance (wave speeds and band gaps) based on the analytical model. The detailed steps are omitted here. A simplified example can be found in the previous work \cite{Morris2022}. Other application examples are discussed below. 

\subsection{Level repulsion identification}\label{sec:lr}

The micro-geometry and periodicity of MMs add an extra layer of complexity to the analysis of band topology and scattering response. The high dimensionality of traditional models presents challenges in understanding physical phenomena and interpreting results. In MM and phononic band structures, many apparent crossing points may exist between eigenfrequency branches. It is important to classify these crossings as either degeneracy points (real crossings) or level repulsions (avoided crossings) \cite{Lu2018}. Despite a small quantitative difference between the two types of crossings, this discrepancy can lead to misunderstandings of modal natures and scattering responses. Level repulsion indicates mixing of modes, caused by the coupling between DOFs, resulting in unexpected energy transfer in scattering analysis. Conversely, real crossings indicate fully decoupled modes \cite{Amirkhizi2018d}.

\begin{figure}[!ht]
	\centering\includegraphics[height=150pt]{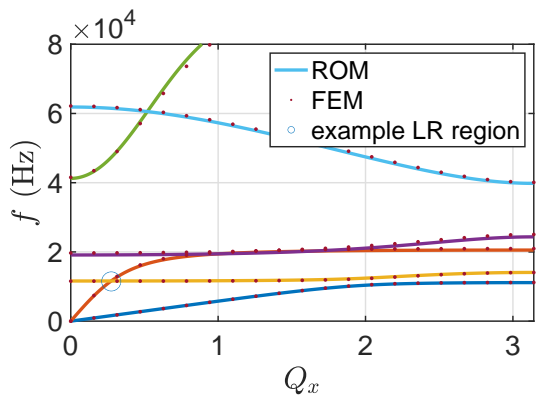}
		\caption{The square MM unit cell band structure with an apparent or real crossing region identified in the blue circle.\label{fig:lr_band}}
\end{figure}

\begin{figure}[!ht]
	\begin{subfigure}[b]{1\linewidth}
		\centering\includegraphics[height=120pt]{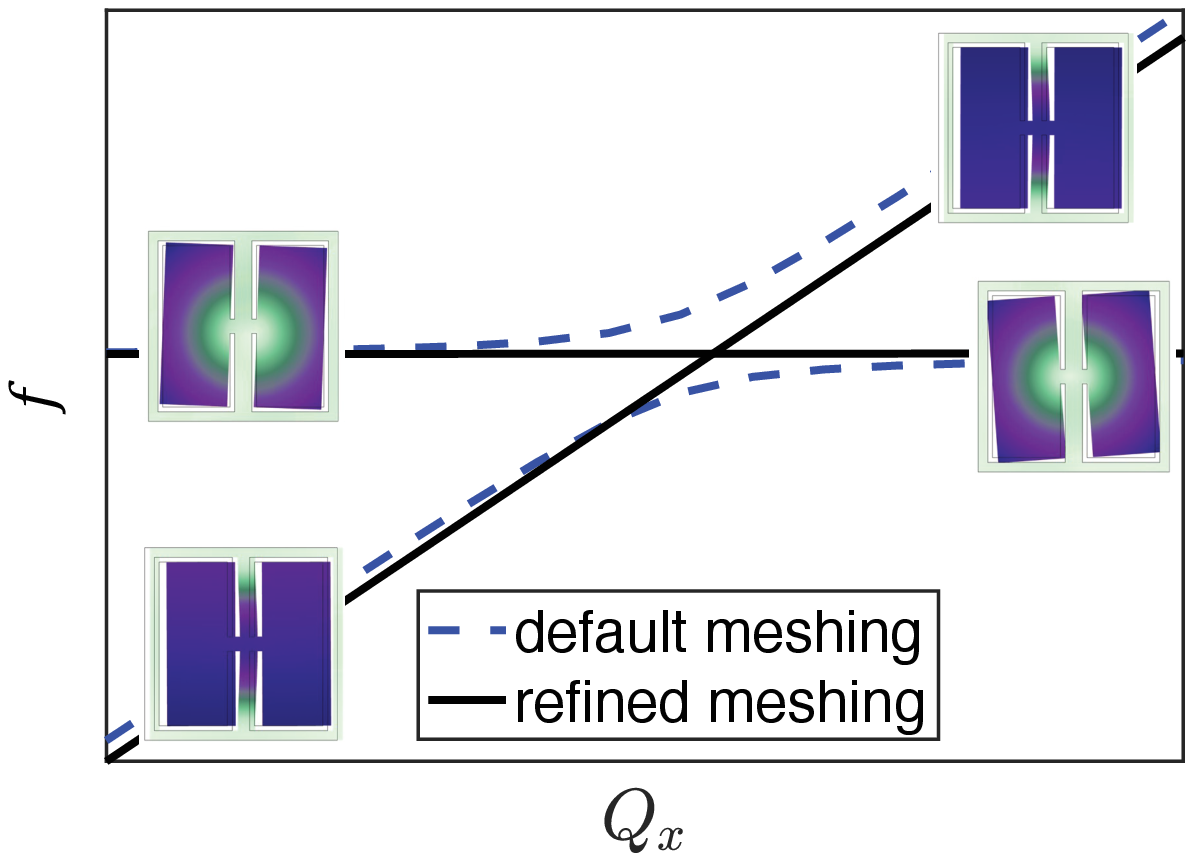}
		\caption{\label{fig:lrfem}}
	\end{subfigure}\\
		\begin{subfigure}[b]{1\linewidth}
		\centering\includegraphics[height=120pt]{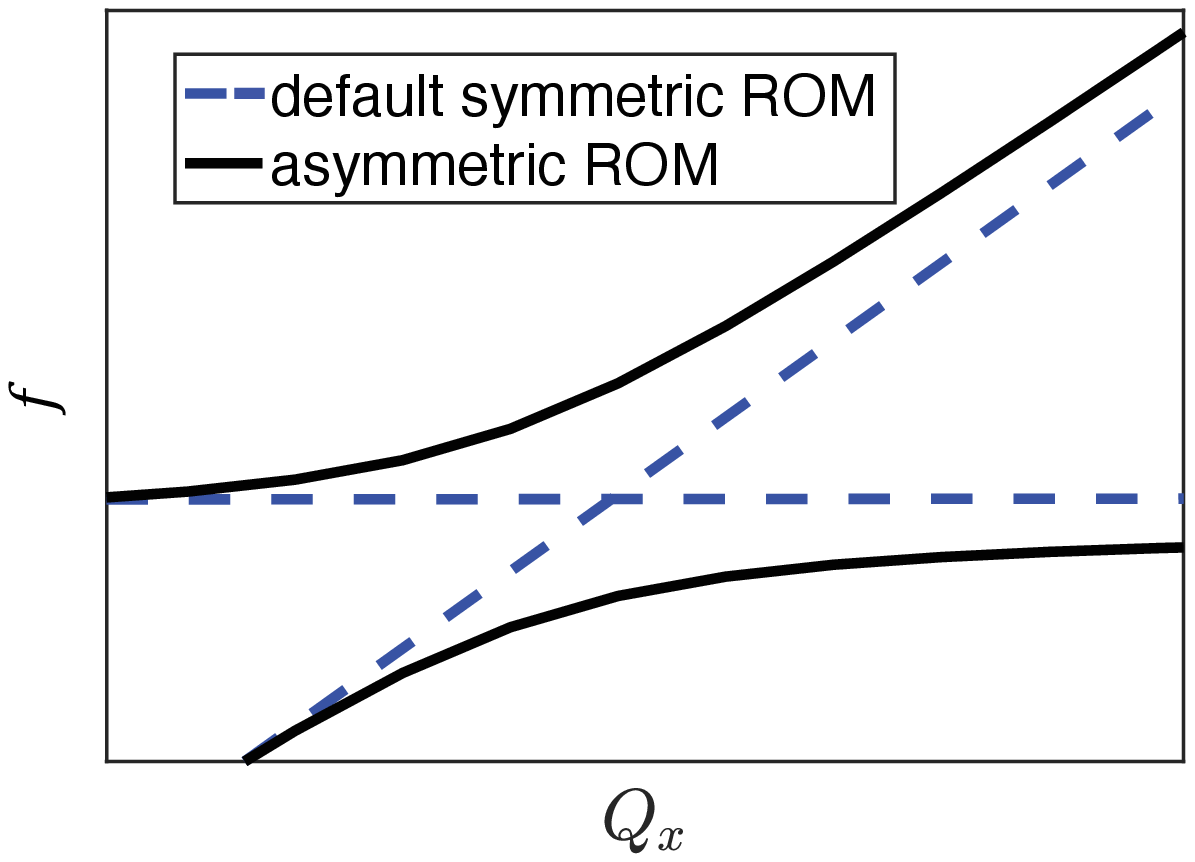}
		\caption{\label{fig:lrrom_band}}
	\end{subfigure}%
	\caption{Band crossing identification example for the square MM cell. The small region identified with a circle in \cref{fig:lr_band} is magnified for both computational approaches: (\subref{fig:lrfem}) FEM, (\subref{fig:lrrom_band}) ROM. In both cases slight baseline shift in the asymmetric band is applied to bring the two results in the same narrow frequency window. \label{fig:crossing}} 
\end{figure}

A demonstration of band identification can be seen in \crefrange{fig:lr_band}{fig:crossing}, where the 1D band structure along the $\Gamma-X$ direction is analyzed for the previously shown square cell. \Cref{fig:lr_band} shows good overall agreement between the ROM and FEM results. However, zooming into a region with an apparent crossing (indicated by the blue circle in \cref{fig:lr_band}) reveals a discrepancy. The FEM results with default meshing, shown as the blue dashed curves in \cref{fig:lrfem}, clearly indicate that the two relevant branches appear repulsed with each other. On the other hand, the ROM results, shown by the blue dashed curves in \cref{fig:lrrom_band}, indicate a real crossing between the two branches. 

To confirm that the real crossing indicated by ROM is a correct observation, we calculate the geometric phase along a prescribed wavenumber path in the complex domain, see \cref{app:geometric_phase} for details. The computed geometric phase is zero, suggesting a real crossing point indeed. It is noticed that the discrete model possesses the same symmetry group as the idealized continuum one, while the FEM model may not have such symmetric properties due to the mesh imperfection.
To further investigate the source of such a discrepancy, a manual perturbation to the parameterized ROM quantities is performed to break the two-fold symmetry of the ROM system. Then, repulsed branches are found in the symmetry-broken ROM band structure, as shown by the black curves in \cref{fig:lrrom_band}, similar to the FEM results with default meshing. In this case, an exact geometric phase of~$\pi$ is accumulated after two loops in the prescribed wavenumber path, indicating the existence of exceptional point \cite{Wang2021exceptional,Lu2018} in the complex domain, which is a known companion of level repulsion. Such an analysis using geometric phase calculations is rather easily implemented with the ROM formulation, but it can be extremely challenging for FEM because of the need for evaluating high-dimensional eigenvectors of large-sized non-Hermitian matrices. As the computational complexity of eigen- problems is of the order~$\mathcal{O}(n^3)$, the cost of ROM is significantly reduced (for both the band computation and the topological invariant computation).

The ROM analysis with perturbation in symmetry reveals the source of discrepancy, i.e., the root of this apparent repulsion is not associated with the cell response, but rather due to asymmetry in the numerical mesh. We then validate this conclusion by adjusting the FEM mesh with enforced two-fold symmetry. The resulting corrected FEM band, shown by the black curves in \cref{fig:lrfem}, exhibits a real crossing point, which is the correct topology for an idealized symmetric model.
This identification task requires repeated simulations and mesh refinements with the FEM approach. On the other hand, the analytical representation of the ROM formulas allows for easy distinction \cite{Amirkhizi2018d,Wang2021exceptional} between real crossings and level repulsions. This analysis shows that the studied crossing point is a symmetry-protected degeneracy (rather than an accidental one), which is unstable and prone to forming repulsed branches due to imperfections in the material or geometry of an actual specimen. The same is true for continuum FE models, which may lack symmetry due to their meshes. 

Correct identification of these band topologies is crucial for understanding the dynamic behaviors in the scattering response, especially with the presence of the potential symmetry breaking that may lead to further misunderstandings.
The ROM approach is more efficient for band sorting purposes (and can easily be leveraged in the calculation of topological invariants based on path integrals), and it provides benefits in understanding the difference between an ideal model and a realistic one, as well as in distinguishing the ``normal'' and ``accidental'' degeneracy.

\subsection{Equi-frequency contours}

The developed ROM matrices also allow for computation of the equi-frequency contours~$k_x(\omega,k_y)$ or~$k_y(\omega,k_x)$, i.e., to find the wavevector solutions~$k_{x}$ or~$k_y$ at prescribed~$k_{y}$ or~$k_x$ and frequency values. For complex $k$ components, the solved eigen-modes are the propagating and evanescent waves that constitute the basis solution for the oblique scattering problem.
Such problems are rarely studied for metamaterials but are of prime importance for analyzing the scattering dynamics \cite{Mokhtari2019a}. Solving this type of problem in FEM is currently impractical (but not impossible), because the real-valued frequency~$\omega$ can not be easily enforced in the traditional eigenfrequency study for given complex $k$ components. It is also generally not straightforward to assign a wavenumber component as an eigenvalue to be found, though for phononic media an elegant mixed eigenvalue approach is presented in \cite{Mokhtari2019a}. For metamaterials with complex internal features, the proposed ROM approach would be an ideal alternative. Using the ROM matrices, this problem can be simply solved by finding the global minima of the determinant~$\mathrm{det}\left[\vec{K}^\mathrm{r}(k_x,k_y)-\omega^2\vec{M}^\mathrm{r}\right]$ in the complex~$(k_x,~k_y)$ space for prescribed~$\omega$ values. Examples of such contour calculations are deferred to focused studies on their use.

\subsection{Finite array transient response}
The band structure computations for infinitely periodic arrays are the main tool and primary product of the ROM approach. However, it is of practical interest to study the dynamic response of finite-sized arrays under localized loading or scattering. With the unit cell matrices obtained from the ROM procedure, it is possible to assemble multiple ROM unit cells to model a finite-sized array, easily removing $\vec{k}$ dependence in the stiffness matrix. This reduced order representation of finite systems allows for very fast computation of the frequency and time domain responses of the structure. Non-uniform and non-periodic arrays can be designed by stacking different unit cells, suitable for design for novel applications \cite{vae_paper} such as clocking and insulation. 
Time domain solutions can be directly calculated through time marching integration schemes \cite{ga_paper}. Here, we discuss the use of Duhamel integral solutions for solving such time domain problems. Other methods such as the micropolar-type model, which is particularly suitable for studying the rotational effects, can be used as well \cite{Schiavone2021}. The governing partial differential equations of such an array, after the modal transformation, will read:
\begin{equation}
\vec{\overline{M}}\vec{\ddot{q}}+\vec{\overline{K}}\vec{q}=\vec{\Phi}^\dagger\vec{F}(t),
\end{equation}
where~$\vec{\overline{M}}=\vec{\Phi}^\dagger\vec{M}\vec{\Phi}$ is the diagonal modal mass matrix,~$\vec{\overline{K}}=\vec{\Phi}^\dagger\vec{K}\vec{\Phi}$ is the diagonal modal stiffness matrix, both associated with and assembled for the full finite array, ~$\vec{\Phi}$ is the eigenvector matrix, also associated with the finite structure, and which needs to be calculated,~$\vec{F}$ is the nodal load vector,~$\vec{q}$ is the generalized coordinate vector, and~$\vec{u}=\vec{\Phi}\vec{q}$ is the nodal DOF vector. Such a form decouples the equations and renders a set of single DOF equations to solve. For a single DOF system with mass~$M$, stiffness~$K$, eigenfrequency~$\omega$, displacement~$u$, and loading~$F(t)$, the dynamic response under arbitrary loading can be computed using the Duhamel integral:
\begin{equation}
    u(t)=\int_0^\tau F(\tau)h(t-\tau)\mathrm{d}\tau
\end{equation}
where the impulse response function (IRF) is
\begin{equation}
    h(t)=\frac{1}{M\omega}\exp[\sin(\omega t)].
\end{equation}

\begin{figure}[!ht]
		\centering\includegraphics[height=130pt]{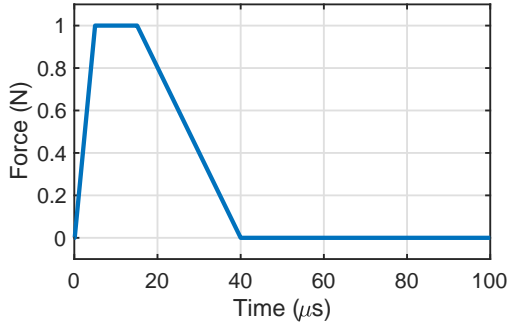}
	\caption{Time domain loading profile.}
	\label{fig:load}
\end{figure}

\begin{figure}[!ht]
		\centering\includegraphics[height=160pt]{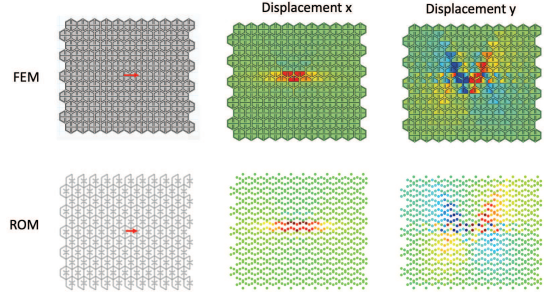}
	\caption{Time domain response of a hexagonal MM array $40\ \mu$s after loading initiation. Top panel: FEM model and solutions. Bottom panel: ROM solutions.}
	\label{fig:td_res}
\end{figure}

Therefore, one can compute each entry in~$\vec{q}(t)$ in a similar way. Then, the physical response will be~$\vec{u}(t)=\vec{\Phi}\vec{q}(t)$. An example of time dependent response computation is shown in \cref{fig:td_res}. An impact force (\cref{fig:load}) is horizontally applied to one of the internal nodes, as indicated by the red arrows in \cref{fig:td_res}. It can be seen that the ROM solution can accurately reproduce the wave propagation pattern. The resulting displacement field from ROM has 94\% R-squared correlation with the FEM data, showing high physical fidelity.
Finally, our preliminary results show that using the time marching approach for the shown system, the FEM model has approximately 700,000 DOFs while the ROM only needs 2,000. Consequently, the computation speed of ROM is about 5,000 times faster than the FEM approach. Therefore, the computational efficiency can be significantly improved for large-sized arrays. Along with data-driven and machine learning methods, one can efficiently explore a vast design space and achieve fast cell-by-cell optimization using the proposed ROM method \cite{vae_paper}. 

It must be noted that the modeling of finite arrays requires an extra step in terms of characterizing the boundary elements. The structural components of the main body remain the same as the ones obtained from the unit cell ROM. However, the outermost elements need to be quantified separately due to the different boundary conditions assigned to them, especially when the MM array is in contact with a different homogeneous medium, or exposed to traction and/or displacement boundary conditions. The detailed approach is subject of current research and is initially carried out by optimizing the static response or impedance of edge cells with the help of a few FEM simulations \cite{ga_paper,josh_skin}. 

\section{Conclusion and Outlook}\label{sec:conclusion}

A general reduced order modeling technique for periodic mechanical metamaterials is introduced. The method uses a limited number of simulations at selected wavevector locations to establish the reduced system matrices, which are parameterized based on the structural connectivity. Effective parameters are extracted by matching modal energies. This approach expands upon previous model order reduction techniques by considering the wave's propagating nature, leading to accurate eigen-solutions for any wavevector. The parameterized and analytical matrices generated by the ROM method offer valuable insights into the micro-structural influence on overall structural behavior and provide significant assistance in fine-tuning of design. Additionally, the ROM approach leads to fast computation of the dynamic responses of finite-sized arrays, with a significant reduction in computational effort compared to FEM.

To summarize, the highlights of this work are: 1) the proposed ROM method can be easily applied to any periodic metamaterial that has beam-like components;
2) the reduced order matrices allow fast and accurate computation of band structures and the dynamic responses in frequency and time domains; 3) the ROM method can further benefit design optimization due to its computational efficiency.

The essential limitation of the proposed work is that the ROM method can only be applied to MM micro-structures comprised of beam-like elements (and potentially plate-like elements). Other types of micro-structures, such as layered media, or unit cells with solid inclusions in a solid matrix, are not the suitable modeling target for the proposed ROM (instead, one can use RBME \cite{Hussein2009a}). 
In addition, the proposed ROM approach is only applicable to systems with stiffness coupling between nearest neighbors, i.e., the long-range interaction \cite{Farzbod2020} is not considered. 

The presented method could contribute to the modeling and design of finite and periodic mechanical metamaterials by reducing the computational effort. The micro-structure and periodicity of MMs lead to exciting dynamic properties and present theoretical questions in the physics of micro-structured media. The ROM method, equipped with the vibration and strength of material domain knowledge, can offer concise descriptions of the micro-structural dynamics and is a solid analytical tool for studying metamaterial dynamics. In addition, the presented method leads to significant improvements in computational efficiency and is a promising candidate for further design optimization of graded MM arrays with data-driven techniques. 

An immediate topic of research is the handling of edge cells due to their different dynamic response, while the interior cells appear to be very well represented by the ROM based on infinitely periodic media. Future work to be implemented is to adjust the element stiffness matrix formulation \cref{eq:beamK} for compatibility with 3D (and potentially composite) beam and plate elements. In addition, the optimization approaches for matching global quantities between the FEM and ROM can be adjusted, so that lossy elements (viscoelastic material) are allowed in the system. 
These future potentials would extend the modeling capability to 3D designs and enlarge the feasible design space. 

In terms of theoretical advancements, it is shown that the ROM representation allows for the efficient identification of band topology, as the geometric phase can be computed with the small-sized matrices and minimal effort. As a future direction, we suggest further truncating the ROM system to a second-order one, to approximate the local topology near those (apparent) crossing regions of the band structure. Then, only the two relevant modes and their associated DOFs are involved. The truncated ROM matrices would lead to a simple yet elegant representation for analytical investigation of the band topology, for example, see \cite{Wang2021exceptional}. It is desired to use the derived mode shapes near such locations as the basis for high-sensitivity detection. 

\begin{acknowledgment}
The authors wish to thank US Army Research Laboratory for continued support throughout this effort. This research was supported by DEVCOM ARL through Cooperative Agreements W911NF-17-2-0173 and W911NF-20-2-0147.
\end{acknowledgment}

\bibliographystyle{asmems4}
\bibliography{asme2e}

\appendix
\section{Beam Element Analysis}\label{appBeam}

\begin{figure}[!ht]
		\centering\includegraphics[height=105pt]{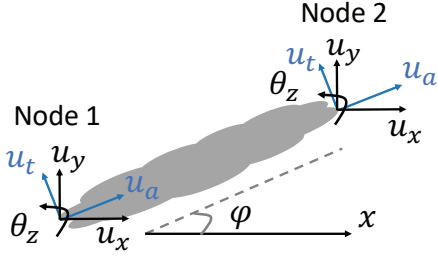}
	\caption{A generalized beam element connecting two nodes with six DOFs. The superscripts (1,2) denoting the nodes are omitted in the figure.  \label{fig:genBeam}}
\end{figure}

The structural connection between two nodes can be modeled as a generalized beam element, as shown in~\cref{fig:genBeam}. This element has two displacement DOFs and one rotation DOF at each end. The beam axial direction makes an angle~$\varphi\in[-\pi/2,\pi/2)$ with the~$x$ axis. The axial DOF~$u_a$ and axial force~$F_{a}$ 
are parallel to the beam axis. The transverse DOF~$u_{t}$ and force~$F_{t}$ are normal to the beam axis. The nodal rotation and applied moment are denoted by~$\theta_z$ and~$M_z$, respectively. In the local coordinate system, the force-displacement equations can be written as
\begin{equation}
    \vec{F}^{\mathrm{local}}=  \vec{K}^{\mathrm{local}}  \vec{U}^{\mathrm{local}},
\end{equation}
i.e.,
\begin{equation}\label{eq:beamK}
    \begin{bmatrix}
    F_{a}^{(1)}\\ F_{t}^{(1)} \\M_z^{(1)}\\
     F_{a}^{(2)}\\ F_{t}^{(2)} \\M_z^{(2)}\\
    \end{bmatrix}=
    \begin{bmatrix}
       \beta_{1}    & 0          & 0          & -\beta_{1} & 0           & 0           \\
              & \beta_{2} & \beta_{3} & 0           & -\beta_{2} & \beta_{4}  \\
              &            & \beta_{5} & 0           & -\beta_{3} & \beta_{6}  \\
              &            &            & \beta_{1}  & 0           & 0           \\
              &            &            &             & \beta_{2}  & -\beta_{4} \\
\mathrm{sym.} &            &            &             &             & \beta_{7} 
    \end{bmatrix}
     \begin{bmatrix}
    u_a^{(1)}\\ u_t^{(1)} \\\theta_z^{(1)}\\
   u_a^{(2)}\\ u_t^{(2)} \\\theta_z^{(2)}\\
    \end{bmatrix},
\end{equation}
where the superscripts denote different nodes.
Here the beam element is allowed to be non-prismatic. The stiffness matrix is always symmetric due to reciprocity. There are only seven different non-zero values~$\beta_{1-7}$ in the stiffness matrix, and they are positive real numbers determined by the element geometry and the material properties. Using machine learning and regression approaches, these spring constants can be related to the detailed geometry and material properties \cite{Morris2022}.

Additional constraints must be imposed on the beam parameters. Moment balance requires that
$$F_{t}^{(1)}L=M_z^{(1)}+M_z^{(2)},$$
where~$L$ is the distance between the two end nodes, and which should be satisfied for any combination of prescribed displacements. Therefore, given the length~$L$, a beam only has four independent parameters~$\beta_{1,2,5,7}>0$, and the other parameters~$\beta_{3,4,6}$ can be determined:

\begin{equation}
\begin{aligned}
\beta_3&=\frac{1}{2L}(\beta_2 L^2+\beta_5-\beta_7)\\
\beta_4&=\frac{1}{2L}(\beta_2 L^2-\beta_5+\beta_7)\\
\beta_6&=\frac{1}{2}(\beta_2 L^2-\beta_5-\beta_7)
\end{aligned}
\end{equation}

To assemble the stiffness matrices of multiple beam elements, it is required to first convert the loads and DOFs into the global coordinate system. Then, the beam equations can be written as
\begin{equation}
    \vec{F}^{\mathrm{global}}=  \vec{R}^{-1}\vec{K}^{\mathrm{local}} \vec{R} \vec{U}^{\mathrm{global}},
\end{equation}
where the rotation matrix is 
\begin{equation}
    \vec{R}=\begin{bmatrix}
        \cos(\varphi) & \sin(\varphi) & 0 & 0 & 0 & 0 \\
         -\sin(\varphi) & \cos(\varphi) & 0 & 0 & 0 & 0 \\
          0 & 0 & 1 & 0 & 0 & 0 \\
          0& 0& 0&  \cos(\varphi) & \sin(\varphi) & 0 \\
           0& 0& 0&   -\sin(\varphi) & \cos(\varphi)  & 0 \\
            0 & 0 & 0 & 0 & 0 & 1 \\
    \end{bmatrix}.
\end{equation}

\section{Geometric phase}\label{app:geometric_phase}

The geometric phase~\cite{Mailybaev2005,Asboth2016} represents the quantification of the changes that the state of an adiabatic system acquires when it traverses along a closed path in its parameter space. The geometric phase is gauge invariant~\cite{Asboth2016}, i.e. different normalizations used at various points of the parameter space do not affect it. Therefore it can characterize the topological properties of the eigenfrequency band structure. Since this approach is originally formulated in quantum mechanics, in the context of mechanical waves, one can first convert the governing equation into a Schr\"{o}dinger-type form. Following a similar approach~\cite{Susstrunk2016}, the wave equation is rewritten as
\begin{equation}
    \vec{H}\vec{\Phi}=-\omega\vec{\Phi},
\end{equation}
where
$$\vec{H}=\im \begin{pmatrix}
\vec{0} & \vec{I}\\
-\vec{\vec{M}^{-1}\vec{K}} & \vec{0}
\end{pmatrix}.$$
The right eigenvector~$\vec{\Phi}$ contains the complex displacement and velocity amplitudes of the RUC. The eigenfrequencies of~$\vec{H}$ are positive and negative square roots of the eigenvalues of~$\vec{K},\,\vec{M}$ system. 
Here we consider a slow cyclic variation of complex wavenumber~$Q=Q_x$, which parameterizes $\vec{H}$. The state may start from a certain eigen-mode and evolve continuously along a closed path in the complex~$Q$ space. The geometric phase picked up after the evolution is defined as
\begin{equation}
    \gamma=\oint \im \vec{\Phi}^{L\dagger}\partial_Q\vec{\Phi} \mathrm{d}Q,
\end{equation}
where~$\vec{\Phi}^{L}$ and~$\vec{\Phi}$ are the left and right eigenvectors of the instantaneous mode along the~$Q$ path, and $\partial_Q$ represents the derivative of the eigenvector along the path.

\begin{figure}[!ht]
	\begin{subfigure}[b]{1\linewidth}
		\centering\includegraphics[height=127pt]{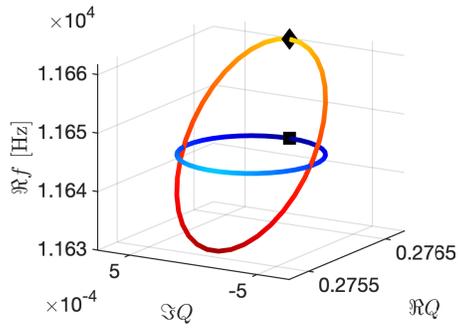}
		\caption{\label{fig:romdg}}
	\end{subfigure}\\
	\begin{subfigure}[b]{1\linewidth}
		\centering\includegraphics[height=127pt]{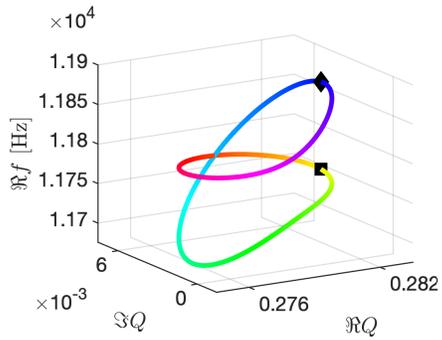}
		\caption{\label{fig:romep}}
	\end{subfigure}
	\caption{ Eigenfrequency trajectories corresponding to closed loops in~$Q$ space. Case (\subref{fig:romdg}) shows the two relevant modes for symmetric ROM with a real crossing in the band. Case (\subref{fig:romep}) shows the two relevant modes for asymmetric ROM with level repulsion in the band.\label{fig:rom_gp}}
\end{figure}

The eigenfrequency trajectories of the parametric evolution in wavenumber~$Q$ are shown in \cref{fig:rom_gp}. We first consider the case of evolution near the real crossing zone with the symmetric ROM setup, as shown in \cref{fig:romdg}. Both two initial states, denoted by~$\blacksquare$ and~$\bLozenge$, are completely recovered after one cycle. The geometric phase picked up by either of the trajectories is exactly zero. No mode-switching behavior can be found. When the paths cross each other in $f$ space, ensuring that there is no discontinuity in the mode shape will select the right choice of path.

For the case of asymmetric ROM with level repulsion, the trajectories are shown in \cref{fig:romep}. The state evolution continuously follows the complex~$f$ trajectory. Due to the existence of an exceptional point \cite{Lu2018} and the Riemann sheet structure of the eigenfrequency surfaces, the eigenmode switches after one cycle instead of recovering back to the original mode, i.e.,~$\blacksquare\rightarrow\bLozenge$ and vice versa. After two cycles, the state goes back to the initial mode with an accumulated extra geometric phase of~$\pi$:~$\{\blacksquare,\bLozenge\}\rightarrow\{-\blacksquare,-\bLozenge\}$. To fully restore the initial starting mode, four cycles will be needed. Similar observations have been made in a micro-cavity experiment in which an EP is encircled~\cite{Dembowski2004}. We refer to Refs.~\cite{Dembowski2004,Mailybaev2005} for detailed theory and Ref.~\cite{Doppler2016} for an experimental study of dynamically encircling an exceptional point.

\end{document}